\documentclass[conference]{IEEEtran}
\usepackage{psfrag}

\usepackage{cite}

\ifCLASSINFOpdf
   \usepackage[pdftex]{graphicx}
\else
   \usepackage[dvips]{graphicx}
\fi
\usepackage{amssymb,amsthm}
\usepackage[cmex10]{amsmath}

\usepackage{algorithm}
\usepackage{algorithmic}

\newcounter{mytempeqncnt}

\begin{document}
%
\title{Analytical Assessment of Coordinated\\ Overlay D2D Communications}
\author{\IEEEauthorblockN{Stelios~Stefanatos,~Antonis~G.~Gotsis,~and~Angeliki~Alexiou,}
\IEEEauthorblockA{Department~of~Digital~Systems,~University~of~Piraeus,~Greece}}


%


\maketitle

\begin{abstract}
In this paper, analytical assessment of overlay-inband device-to-device (D2D) communications is investigated, under cellular-network-assisted (coordinated) scheduling. To this end, a simple scheduling scheme is assumed that takes into account only local (per cell) topological information of the D2D links. Stochastic geometry tools are utilized in order to obtain analytical expressions for the interferers density as well as the D2D link signal-to-interference-ratio distribution. The analytical results accuracy is validated by comparison with simulations. In addition, the analytical expressions are employed for efficiently optimizing the parameters of a cellular system with overlay D2D communications. It is shown that coordinated scheduling of D2D transmissions enhances system performance both in terms of average user rate as well as maximum allowable D2D link distance.
\end{abstract}



%
\IEEEpeerreviewmaketitle

\section{Introduction}
Device-to-Device (D2D) communications have recently attracted a lot of attention as a means to enhance cellular network performance by direct communication between physically proximal cellular devices \cite{Doppler}. However, their incorporation introduces challenges to system design as there are issues related to interference management and sharing of system resources. Although D2D transmissions can be viewed as a degenerated case of an ad-hoc network, therefore allowing for the incorporation of the many techniques already proposed in that field of research, this view neglects the availability of cellular infrastructure that can be utilized to more efficiently perform tasks such as establishment of D2D links and resource allocation \cite{Fodor}.

There has been much research related to (network-assisted) D2D communications, usually by formulation of appropriate optimization problems (see \cite{Asadi} for a recent literature review). However, this approach does not lead to analytical results on the performance of D2D communications, which are of interest as they provide quantitative insights on the benefits of D2D communications as well as guidelines on related aspects of system design. To this end, tools from stochastic geometry \cite{Baccelli} that have been successfully applied previously for ad-hoc and cellular networks analysis have been recently incorporated in D2D-related studies.

Analytical evaluation of D2D communications under this framework is investigated in \cite{Erturk, Lee, Lin} for various system model assumptions, e.g., overlay/underlay D2D communications, power control, e.t.c. However, the issue of resource management (and its potential benefit) has not been addressed, i.e., D2D transmissions are treated as an uncoordinated ad-hoc network where interferers that are in (very) close proximity may exist. In \cite{Huang, Ye}, this deficiency is partially overcome by assuming a random time-frequency hopping channel access scheme for D2D transmissions. This approach improves performance as the probability of having close-by interferers on the same subchannel is reduced. However, it is expected that a more intelligent scheduling scheme, assisted by the cellular infrastructure, will provide better performance.

In this paper, analytical evaluation of overlay-inband D2D communications with network-assisted (coordinated) scheduling is investigated. To this end, a simple scheduling scheme is assumed that takes into account only local (per cell) topological information of the D2D links. Stochastic geometry tools are utilized in order to obtain analytical expressions for the interferers density as well as the D2D link signal-to-interference-ratio (SIR) distribution. The resulting integral-form expression for the SIR distribution can be easily evaluated numerically and provides very good accuracy as evident by comparison with simulation results. In addition, it allows to efficiently perform design optimization tasks and offers an example case study, showing that, under optimized system parameters, availability of D2D communications with coordinated scheduling enhances system performance by increasing the average user rate as well as the maximum allowable D2D link distance.

The paper is organized as follows. Section II describes the system model as well as scheduling schemes for D2D transmissions. In Section III, an analytical expression for the SIR distribution under coordinated scheduling is derived. Section IV provides numerical results validating the accuracy of the analysis and a case study example of the benefits of (coordinated) D2D communications as overlay in the downlink of a cellular network. Section V concludes the paper.

\emph{Notation:} $\mathbb{R}$, $\mathbb{N}$ denote the sets of real and integer numbers, respectively. $\mathbb{P}(\cdot)$ denotes the probability measure, $\mathbb{E}(\cdot)$ the expectation operator, written as $\mathbb{E}(\cdot|\mathcal{E})$ when conditioned on the event $\mathcal{E}$, $|x|$ is the norm of $x \in \mathbb{R}^2$ and $|\mathcal{A}|$ is the area of $\mathcal{A} \subset \mathbb{R}^2$. $B(x,r)$ denotes the ball centered at $x \in \mathbb{R}^2$ with radius $r$, $\overline{\mathcal{A}}$ denotes the complement of $\mathcal{A}$ with respect to (w.r.t) $\mathbb{R}^2$, and $\varnothing$ is the null set. The indicator function is denoted by $\mathbb{I}(\cdot)$, and $\lfloor \cdot \rfloor$ is the integer floor operator.

\section{System Model and Scheduling Schemes}
\subsection{System Model}
A hybrid network is considered that consists of both cellular and D2D links. The locations of the access points (APs) are modeled as a homogeneous Poisson point process (PPP) $\Phi_a = \{\tilde{x}_i\} \subset \mathbb{R}^2$ of density $\lambda_a$. In order to avoid intersystem interference, an overlay-inband D2D communications scheme is assumed, where D2D transmissions are performed on an exclusively assigned part of the available spectrum. The D2D transmitters (TXs) are distributed according to a homogeneous PPP $\Phi_d =\{x_i\} \subset \mathbb{R}^2$ of density $\lambda_d$, independent of $\Phi_a$. Each D2D TX is associated with a unique receiver (RX) that is located at a fixed (worst-case) distance $r_d$ away with isotropic direction. Random link distances for the D2D links can be easily incorporated in  the considered framework by an extra averaging operation. 

All nodes in the system are equipped with a single antenna, with TXs having full buffers and RXs treating interference as noise, i.e., no sophisticated decoding algorithms are considered. All D2D TXs transmit with the same fixed power, normalized to unity, and (thermal) noise power at RXs is assumed negligible compared to interference. 

\subsection{Scheduling Schemes}
In order to control the interference level, the dedicated bandwidth for D2D transmissions is partitioned into $N$ subchannels (SCs) of equal size and each D2D TX transmits in only one of them (the possibility of one SC used by many D2D links is allowed). Clearly, the SC allocation scheme, which, in turn, will affect the choice of $N$, is of critical importance. An uncoordinated (probabilistic) scheduling scheme is the simplest option, defined as follows.
\theoremstyle{definition}
\newtheorem{defn}{Definition}
\begin{defn}
\emph{Uncoordinated scheduling}: Each D2D TX randomly and independently selects one of the $N$ available SCs for transmission.
\end{defn}
Note that under uncoordinated scheduling there are no guarantees on the interference level as it is possible to have many close-by D2D TXs using the same SC. Intuitively, the probability of this event can be made (arbitrarily) small by increasing $N$, however, with the cost of reduced bandwidth utilization. An alternative option is to exploit the cellular infrastructure in order to coordinate D2D transmissions that are in close proximity. Defining the (Voronoi) cell of an AP $\tilde{x}_i$ as the set $\mathcal{C}_i \triangleq \{x \in \mathbb{R}^2 : \|x-\tilde{x}_i\| \leq \|x-\tilde{x}_j\|, j \neq i\}$, the coordinated scheduling scheme considered in this paper is the following.

\theoremstyle{definition}
\newtheorem{def2}[defn]{Definition}
\begin{def2}
\emph{Coordinated scheduling}: Each AP $\tilde{x}_i$ independently allocates SCs to the $K_i$ D2D TXs located within $\mathcal{C}_i$ by the following procedure.
\begin{enumerate}
\item{The $K_i$ D2D TXs are randomly partitioned into $\lfloor K_i/N \rfloor$ groups of size $N$ and one group of size $K_i-\lfloor K/N \rfloor N$ (if $K_i>\lfloor K/N \rfloor N$). Members of each group are randomly ordered.}
\item{SC allocation for each group is performed serially, with the $k$-th D2D TX of the group assigned a unique, randomly selected SC out of the set of SCs not previously allocated to one the first $k-1$ D2D TXs of the group.}
\end{enumerate}
\end{def2}
Coordinated scheduling provides soft guarantees on the interference level as it results in mutual orthogonal transmissions for D2D TXs \emph{of the same group}, although it is possible that the same SC may be assigned to more than one D2D TXs belonging to different groups. Specifically, given $K_i$, $N$, each SC will be allocated to either $\lfloor K_i/N \rfloor$ or $\lfloor K_i/N \rfloor +1$ D2D TXs. However, a (small) possibility of having excessively many nearby D2D TXs using the same SC still exists, as it may happen that close-by TXs located near the common edge of adjacent cells are assigned the same SC.

\section{SIR Analysis}
In this section, the SIR distribution under the aforementioned scheduling schemes is investigated analytically. For the analysis, $\Phi_d$ is conditioned on including a particular (typical) TX $x_0$ positioned, without loss of generality, at a distance $r_d=|x_0|$ from the origin, where its intended (typical) RX resides. Note that, by the properties of the PPP, this conditioning is equivalent to adding the point $x_0$ to $\Phi_d$, i.e., $\Phi_d \setminus \{x_0\}$, conditioned on the existence of $x_0$, is distributed as the original, non-conditioned process $\Phi_d$ \cite{Baccelli}. Under the assumptions of Sec. II, the signal-to-interference-ratio (SIR) at the typical Rx on the SC used by the typical D2D link is 

\begin{equation} \label{eq:SIR}
\textrm{SIR} = \frac{g_{x_0} r_d^{-\alpha}}{I},
\end{equation}
with $I \triangleq \sum_{{x}_i \in \hat{\Phi}_d\setminus \{x_0\}}  g_{x_i} |{x}_i|^{-a}$, where $\{g_{x_i}\}$ are the mutually independent fading coefficients between TXs $\{x_i\}$ and the typical receiver at the origin whose marginal distribution is exponential of unit mean (Rayleigh fading), and $\alpha > 2$ is the path loss exponent. The point process $\hat{\Phi}_d \subseteq \Phi_d$ corresponds to the D2D TXs using the same SC as the typical TX and depends on the scheduling scheme. Note that by the symmetry of the system model, (\ref{eq:SIR}) holds for any SC that the typical D2D TX transmits on, which will be assumed in the following to be the first.

\subsection{Uncoordinated Scheduling}
Under uncoordinated scheduling, $\hat{\Phi}_d$ is the point process resulting from independent thinning of $\Phi_d$ with a retention probability $1/N$, therefore, it is a homogeneous PPP with density $\lambda_d/N$ \cite{Baccelli}. In this case, (\ref{eq:SIR}) corresponds to the SIR of the well-studied bipolar ad-hoc network with Poisson distributed interferers whose distribution is given in the following lemma \cite{Muhlethaler,Baccelli}.
\theoremstyle{plain}
\newtheorem{lem}{Lemma}
\begin{lem}
The complementary cumulative distribution function (ccdf) of the SIR with uncoordinated scheduling is
\begin{equation} \label{eq:adhoc_cov}
\mathbb{P}(\textrm{\emph{SIR}} \geq \theta) =  \exp \left(-\frac{\lambda_d}{N} \kappa r_d^2 \theta^{2/\alpha}\right),   \theta \geq 0,
\end{equation}
where $\kappa \triangleq (2\pi^2/\alpha)/\sin(2\pi/\alpha)$.
\end{lem}
Equation (\ref{eq:adhoc_cov}) verifies the intuition that increasing $N$ improves SIR due to a decrease of the number of interferers on the considered SC.

\subsection{Coordinated Scheduling}
It is clear that there is no difference between coordinated and uncoordinated scheduling for $N=1$, therefore, it will be assumed in the following that $N\geq2$. In order to obtain the SIR distribution under coordinated scheduling it is necessary to obtain the statistics of the interference power $I$ or, equivalently, its Laplace transform $\mathcal{L}_I(s) \triangleq \mathbb{E}(e^{-sI})$. To this end, the following assumption is employed that significantly simplifies the analysis.
\theoremstyle{plain}
\newtheorem{as}{Assumption}
\begin{as}
Each D2D RX is located within the same cell as its corresponding D2D TX.
\end{as}
Clearly, this assumption is valid with high probability when $r_d$ is sufficiently small. Let $\tilde{x}_0$ denote the AP closest to the typical D2D link. Then, assumption 1 allows to express the interference power \emph{conditioned on} $\mathcal{C}_0$, $I_{|\mathcal{C}_0}$, as 
\begin{equation} \label{eq:I2}
I_{|\mathcal{C}_0} = I_{\mathcal{C}_0} +  I_{\overline{\mathcal{C}}_0},
\end{equation}
where $I_{\mathcal{C}_0}$ denotes the intracell interference power generated by the point process $\hat{\Phi}_{d,\mathcal{C}_0} \triangleq (\hat{\Phi}_d\cap\mathcal{C}_0)\setminus \{x_0\}$ and $I_{\overline{\mathcal{C}}_0}$ denotes the intercell interference power generated by the process $\hat{\Phi}_{d,\overline{\mathcal{C}}_0} \triangleq \hat{\Phi}_d\cap\overline{\mathcal{C}}_0$. By the properties of the PPP, $(\Phi_d\cap\mathcal{C}_0)\setminus \{x_0\}$ and $\Phi_d\cap\mathcal{C}_0$ are independent PPPs. Since coordinated scheduling decisions within $\mathcal{C}_0$ are independent of decisions in other cells, it follows that
\theoremstyle{plain}
\newtheorem{lem1b}[lem]{Lemma}
\begin{lem1b}
$\hat{\Phi}_{d,\mathcal{C}_0}$ and $\hat{\Phi}_{d,\overline{\mathcal{C}}_0}$ are independent.
\end{lem1b}


\begin{figure*}[!t]
\normalsize
\setcounter{mytempeqncnt}{\value{equation}}
\setcounter{equation}{8}
\begin{equation} \label{eq:cov_prob_cond}
\mathbb{P}(\textrm{{SIR}} \geq \theta|r_a) = \exp\left(-\frac{\lambda_d}{N} \kappa r_d^2 \theta^{2/\alpha}+\frac{\lambda_d}{N}\frac{\Gamma(N-1,\lambda_d|\mathcal{C}_0(r_a)|)}{(N-2)!}\int_{-\phi_0}^{\phi_0}\int_{0}^{r_0(r_a,\phi)}\frac{u}{1+\theta^{-1}(u/r_d)^{\alpha}}dud\phi\right), \theta \geq 0
\end{equation}
\setcounter{equation}{\value{mytempeqncnt}}
\hrulefill
\vspace*{4pt}
\end{figure*}

However, in contrast to the uncoordinated case, $\hat{\Phi}_{d,\mathcal{C}_0}$ and $\hat{\Phi}_{d,\overline{\mathcal{C}}_0}$ are not the result of independent thinning of  $(\Phi_d\cap\mathcal{C}_0)\setminus \{x_0\}$ and $\Phi_d\cap\mathcal{C}_0$, respectively, since the coordinated scheduling process introduces correlation among SC allocations to the D2D TXs within each cell. Therefore, they cannot be claimed to be PPP and their actual distribution must be derived, which is a difficult task. On the other hand, the semi-random manner under which coordinated scheduling is performed suggests that this correlation is small and, therefore, the following approximation is expected to be accurate.
\newtheorem{as2}[as]{Assumption}
\begin{as2}
$\hat{\Phi}_{d,\mathcal{C}_0}$ and $\hat{\Phi}_{d,\overline{\mathcal{C}}_0}$ are homogeneous PPPs.
\end{as2}
The term homogeneous in the above assumption is understood as referring only to the subset of $\mathbb{R}^2$ where the processes are not identically $\varnothing$, i.e., $\hat{\Phi}_{d,\mathcal{C}_0}$ has a constant, non-zero density only in $\mathcal{C}_0$ and zero density in $\overline{\mathcal{C}}_0$, and similarly for $\hat{\Phi}_{d,\overline{\mathcal{C}}_0}$. Assumption 2 is critical as it allows to incorporate well known analytical tools for PPPs, which, however, require knowledge of the densities of $\hat{\Phi}_{d,\mathcal{C}_0}$, $\hat{\Phi}_{d,\overline{\mathcal{C}}_0}$, provided in the following.

\theoremstyle{plain}
\newtheorem{prop}{Proposition}
\begin{prop}
Under assumption 2, the density $\hat{\lambda}_{d,\overline{\mathcal{C}}_0}$ of $\hat{\Phi}_{d,\overline{\mathcal{C}}_0}$ equals
\begin{equation} \label{eq:l_out}
\hat{\lambda}_{d,\overline{\mathcal{C}}_0} =  \frac{\lambda_d}{N},
\end{equation}
and the density $\hat{\lambda}_{d,\mathcal{C}_0}$ of $\hat{\Phi}_{d,\mathcal{C}_0}$ equals
\begin{equation}  \label{eq:l_in}
\hat{\lambda}_{d,\mathcal{C}_0} = \frac{\lambda_d}{N}\left(1-\frac{\Gamma(N-1,\lambda_d|\mathcal{C}_0|)}{(N-2)!}\right),
\end{equation}
where $\Gamma(a,z) \triangleq \int_z^\infty t^{a-1}e^{-t}dt$ is the (upper) incomplete gamma function.
\end{prop}
\begin{IEEEproof}
Recall that the density of a homogeneous point process equals the average number of points within any bounded subset of $\mathbb{R}^2$ divided by its area \cite{Baccelli}. Consider \emph{any} cell $\mathcal{C}_i$, $i\neq 0$, with $K_i$ D2D TXs located within it, and let $\hat{K}_i$ denote the number of D2D TXs assigned SC 1. Noting that, with probability 1, $0 < |\mathcal{C}_i| < \infty, \forall i$, \cite{Baccelli}
\setlength{\arraycolsep}{0.3em}
\begin{eqnarray} \label{eq:proof_out}
\hat{\lambda}_{d,\overline{\mathcal{C}}_0} &{=}& \mathbb{E}(\hat{K}_i)/|\mathcal{C}_i|\nonumber\\
{}&{=}& \mathbb{E}[\mathbb{E}(\hat{K}_i|K_i)]/|\mathcal{C}_i|\nonumber\\
{}&{=}& \mathbb{E}(K_i)/(N|\mathcal{C}_i|),
\end{eqnarray}
where the last equation is obtained by noting that $\mathbb{E}(\hat{K}_i|K_i) = K_i/N$ that can be verified by straightforward computations. Since $\mathbb{E}(K_i) = \lambda_d |\mathcal{C}_i|$, (\ref{eq:l_out}) follows. Now consider cell $\mathcal{C}_0$ and let $K_0$ denote the number of D2D TXs located within $\mathcal{C}_0$ \emph{in addition to} the typical TX, and $\hat{K}_0$ the number of D2D TXs assigned SC 1. Then,
\setlength{\arraycolsep}{0.3em}
\begin{eqnarray} \label{eq:proof_in}
\hat{\lambda}_{d,\mathcal{C}_0} &{=}& \mathbb{E}(\hat{K}_0)/|\mathcal{C}_0|\nonumber\\
{}&{=}& \mathbb{E}[\mathbb{E}(\hat{K}_0|K_0)]/|\mathcal{C}_0|\nonumber\\
{}&\overset{(a)}{=}& \mathbb{E}[(K_0/N)\mathbb{I}(K_0 \geq N)]/|\mathcal{C}_0|\nonumber\\
{}&{=}& \frac{1}{N|\mathcal{C}_0|}\sum_{k=N}^{\infty}k \mathbb{P}(K_0=k)\nonumber\\
{}&\overset{(b)}{=}& \frac{\lambda_d}{N} e^{-\lambda_d|\mathcal{C}_0|}  \sum_{k=N}^{\infty}\frac{(\lambda_d|\mathcal{C}_0|)^{k-1}}{(k-1)!},
\end{eqnarray}
where (a) follows by noting that $\hat{K}_0 = 0$ for $K_0 \leq N-1$, and (b) since $K_0$ is a Poisson random variable of mean $\lambda_d|\mathcal{C}_0|$. Employing the series representation of $\Gamma(n,z)$ for $n \in \mathbb{N}$ \cite[Eq. 8.352.7]{Gradshteyn} results in (\ref{eq:l_in}).
\end{IEEEproof}
Proposition 1 provides some initial insights on the interference levels provided by coordinated scheduling in terms of density of interferers. In particular, the interferers density outside $\mathcal{C}_0$ is the same as in the uncoordinated case, i.e., coordination does not provide any benefit in this respect, which is not surprising as scheduling decisions are taken independently per cell. On the other hand, coordination manages to reduce the density of intra-cell interferers as it can be directly verified that $\hat{\lambda}_{d,\mathcal{C}_0} < \hat{\lambda}_{d,\overline{\mathcal{C}}_0}$. Note also that $\hat{\lambda}_{d,\mathcal{C}_0}$ is a monotonically increasing function of $\lambda_d|\mathcal{C}_0|$, with $\hat{\lambda}_{d,\mathcal{C}_0}\rightarrow \lambda_d/N$ for $\lambda_d|\mathcal{C}_0|\rightarrow \infty$, i.e., \emph{for a fixed value of} $N$, the benefit of coordinated scheduling diminishes as the expected number of D2D TXs within $\mathcal{C}_0$ increases.

Having specified the above densities, the Laplace transform of $I_{\mathcal{C}_0}$  and  $I_{\overline{\mathcal{C}}_0}$ is provided in the following lemma \cite{Baccelli}.
\newtheorem{lem2}[lem]{Lemma}
\begin{lem2}
The Laplace transform of $I_{\mathcal{C}_0}$ is
\begin{equation} \label{eq:L_I_in}
\mathcal{L}_{I_{\mathcal{C}_0}}(s)=\exp\left(-\hat{\lambda}_{d,\mathcal{C}_0} \int_{\mathcal{C}_0}\frac{s}{s+\|x\|^{\alpha}}dx\right).
\end{equation}
The Laplace transform of $I_{\overline{\mathcal{C}}_0}$ is similar to (\ref{eq:L_I_in}) with $\hat{\lambda}_{d,\overline{\mathcal{C}}_0}$ and $\overline{\mathcal{C}}_0$ in place of $\hat{\lambda}_{d,\mathcal{C}_0}$ and $\mathcal{C}_0$, respectively.
\end{lem2}
Unfortunately, $\mathcal{C}_0$ is a convex polyhedron \cite{Baccelli} which makes even numerical evaluation of (\ref{eq:L_I_in}) impractical. Clearly, a simple approximation of $\mathcal{C}_0$ is required and it is natural to consider a circular shape area whose radius depends on the distance $r_a$ between the typical D2D RX and AP $\tilde{x}_0$. In particular, the following two approximations may be employed for $\mathcal{C}_0$.
\theoremstyle{plain}
\newtheorem{as3}[as]{Assumption}
\begin{as3}
$\mathcal{C}_0$ is a circular region equal to (see Fig. 1)
\begin{enumerate}
\item{$\mathcal{C}_0 = \mathcal{B}_1 \triangleq B((0,r_a/2),r_a/2)$, or}
\item{$\mathcal{C}_0 = \mathcal{B}_2 \triangleq B((0,0),r_a)$}
\end{enumerate}
\end{as3}
Note that, by the rotational invariance of the PPPs, $\tilde{x}_0$ has been considered to lie on the x-axis for simplicity \cite{Haenggi}. $\mathcal{B}_1$ is chosen as a conservative approximation of $\mathcal{C}_0$, since $|\mathcal{B}_1| < |\mathcal{C}_0|$ \cite{Akoum} and the typical D2D is placed right on its edge. Approximation $\mathcal{B}_2$ attempts to remedy the latter issue by covering the typical D2D TX from all sides. It can be easily seen that neither approximation will be a good fit for all realizations of $\Phi_a$. However, they do allow for a computationally tractable expression of the SIR distribution under coordinated scheduling, conditioned on $\mathcal{C}_0$, or, equivalently, on $r_a$.

\begin{figure}
\centering
\resizebox{7cm}{!}{\includegraphics[trim = 15mm 5mm 10mm 10mm]{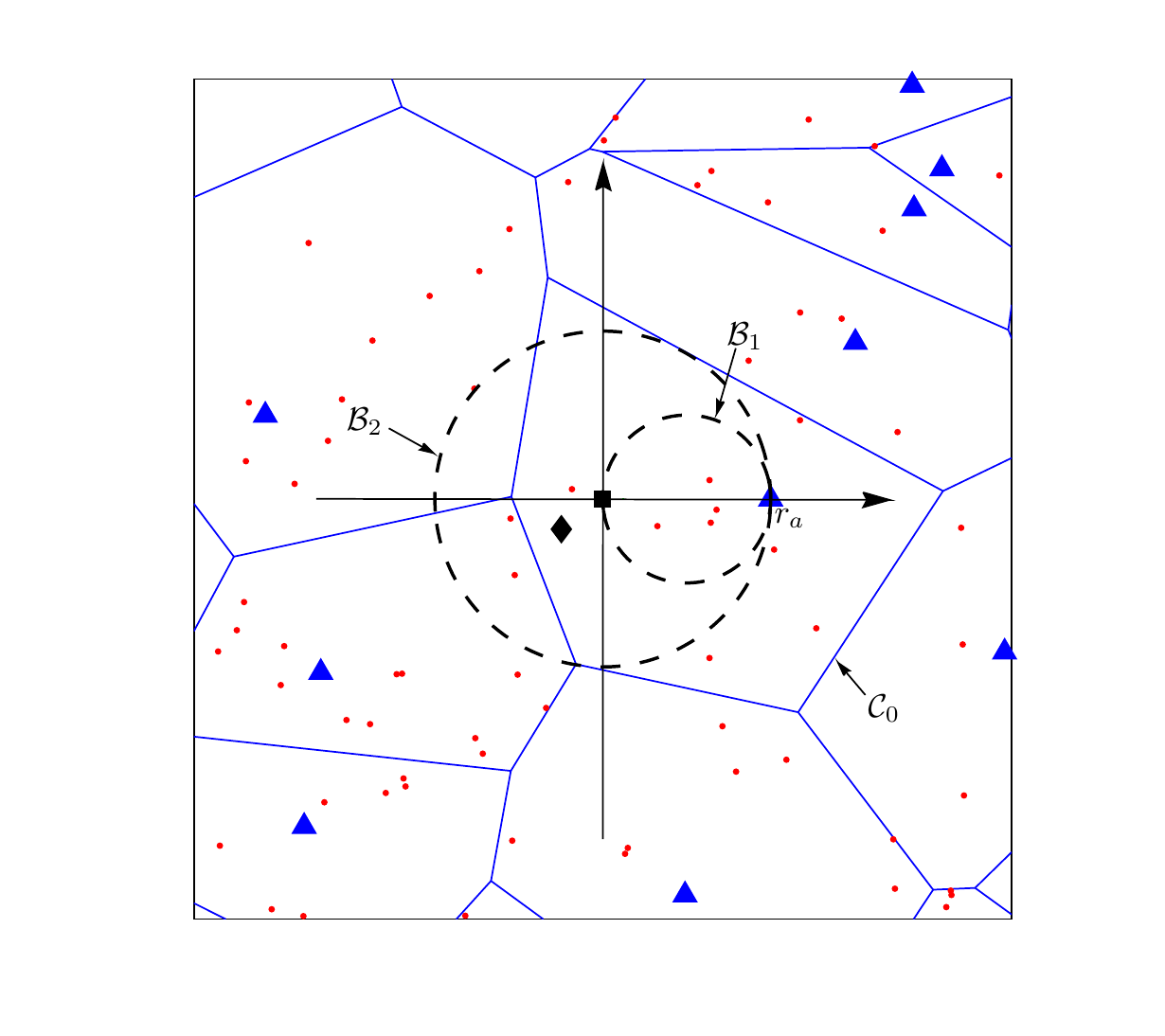}}
\caption{Approximations of $\mathcal{C}_0$. Triangles, dots, square and diamond represent APs, D2D TXs, typical D2D RX, and typical D2D TX, respectively.}
\end{figure}

\theoremstyle{plain}
\newtheorem{prop2}[prop]{Proposition}
\begin{prop2}
Under Assumptions 1-3, the ccdf of SIR with coordinated scheduling and conditioned on the distance $r_a < \infty$ is given by (9) (shown at the top of the page), where $|\mathcal{C}_0(r_a)|=\pi (r_a/2)^2$, $\phi_0=\pi/2$, $r_0(r_a,\phi)=2r_a\cos(\phi)$, under the approximation $\mathcal{C}_0 = \mathcal{B}_1$, and $|\mathcal{C}_0(r_a)|=\pi r_a^2$, $\phi_0=\pi$, $r_0(r_a,\phi)=r_a$, under the approximation $\mathcal{C}_0 = \mathcal{B}_2$.
\end{prop2}
\begin{IEEEproof} \addtocounter{equation}{1}
Starting from (\ref{eq:SIR}),
\begin{eqnarray} \label{eq:proof_cond}
\mathbb{P}(\textrm{{SIR}} \geq \theta|r_a) &{=}&\mathbb{P}(g_{x_0} \geq \theta r_d^{\alpha} I|r_a)\nonumber\\
{}&{=}& \mathcal{L}_{I_{|\mathcal{C}_0}}(\theta r_d^{\alpha})\nonumber\\
{}&{=}& \mathcal{L}_{I_{\mathcal{C}_0}}(\theta r_d^{\alpha}) \mathcal{L}_{I_{\overline{\mathcal{C}}_0}}(\theta r_d^{\alpha}),
\end{eqnarray}
where the last equation follows from (\ref{eq:I2}) and Lemma 2. Using the Laplace transform formula of Lemma 3, replacing $\mathcal{C}_0$ with one of the approximations in Assumption 3 and writing the integrals in polar coordinates results in (9).
\end{IEEEproof}
The integral term of (9) is in a form that allows for numerical computation and can also be written in closed form for the case of $\alpha=4$. In addition, it can be easily seen that it is non-negative for any $r_a$, which, after comparing with (\ref{eq:adhoc_cov}), shows that the SIR under coordinated scheduling stochastically dominates SIR under uncoordinated scheduling, i.e., coordinated scheduling provides at least as good performance as uncoordinated scheduling in terms of SIR.

In order to obtain an $r_a$-independent SIR distribution the statistics of $r_a$ must be determined. For small $r_d$, $r_a$ is close to the distance between $x_0$ and $\tilde{x}_0$ whose distribution is known \cite{Haenggi2}, suggesting the following approximation.
\newtheorem{as4}[as]{Assumption}
\begin{as4}
Distance $r_a$ is Rayleigh distributed with $\mathbb{E}(r_a) = 1/(2\sqrt{\lambda_a})$.
\end{as4}
The unconditioned SIR distribution can now be obtained by simple averaging.
\theoremstyle{plain}
\newtheorem{cor2}{Corollary}
\begin{cor2}
Under assumptions 1-4, the ccdf of SIR with coordinated scheduling equals
\begin{equation}
\mathbb{P}(\textrm{\emph{SIR}} \geq \theta)=\int_0^\infty2\pi \lambda_a r_a e^{-\pi \lambda_a r_a^2} \mathbb{P}(\textrm{\emph{SIR}} \geq \theta|r_a)dr_a,
\end{equation}
with $\mathbb{P}(\textrm{\emph{SIR}} \geq \theta|r_a)$ as in (9).
\end{cor2}

\begin{figure}
\centering
\resizebox{8.5cm}{!}{\includegraphics[trim = 10mm 5mm 10mm 10mm]{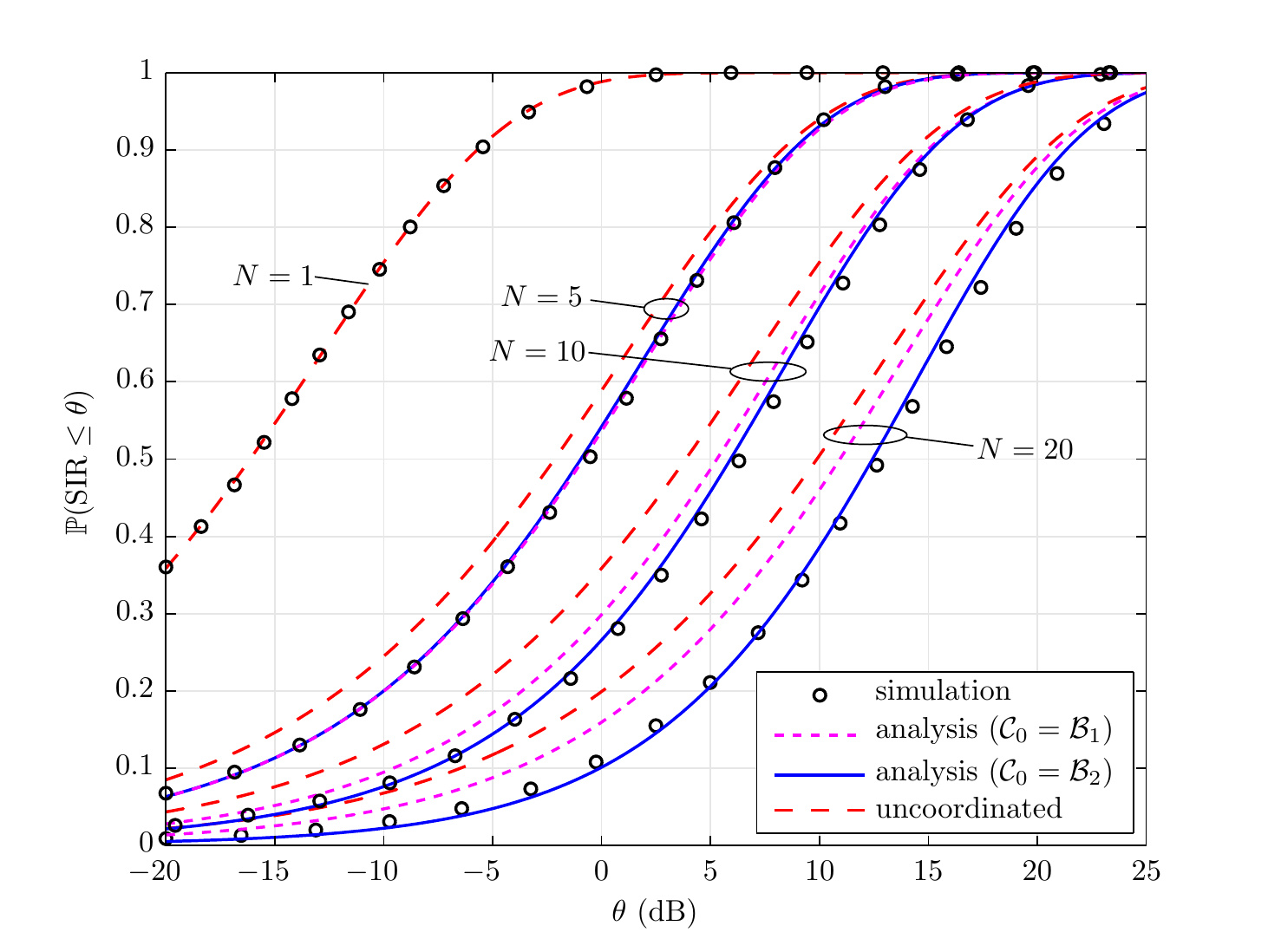}}
\caption{SIR distribution of typical D2D link ($\alpha = 4$, $\lambda_a=1$, $\lambda_d=10$, $r_d = 0.6/(2\sqrt{\lambda_a})$).}
\end{figure}

\section{Numerical Results}
In order to assess the accuracy of the analytical results for the coordinated case, simulations were performed assuming $\lambda_a=1$, $\lambda_d=10$, which corresponds to an average of 10 D2D TXs (links) per cell. In each simulation run, a realization of $\Phi_a$ and $\Phi_d$ was generated in an area of size large enough to include 30 APs on average and the SIR of the typical link was measured. The results depicted were obtained by averaging over $10^6$ independent runs with the value of $\alpha$ set to $4$.

\subsection{SIR Distribution}
Figure 2 shows the cumulative distribution function (cdf) of SIR under coordinated scheduling with $r_d=0.6 / (2\sqrt{\lambda_a})$ (recall that the average distance between typical D2D TX and closest AP is $1/(2\sqrt{\lambda_a})$). Note that there are on average $\lambda_d \pi r_d^2 \approx 2.8$ D2D TXs located closer to the typical D2D RX than the typical D2D TX. Performance for various $N$ is depicted, as obtained by simulations as well as the analytical expression of Corollary 1. In addition, the performance of uncoordinated scheduling (Lemma 1) is also shown without corresponding simulation results as (\ref{eq:adhoc_cov}) is exact. It can be seen that for $N=1$, performance is extremely poor due to  large interference, whereas $N>1$ provides significant performance improvement. The analytical expression for the SIR is very close to the simulation results under the approximation $\mathcal{C}_0 = \mathcal{B}_2$, whereas approximation  $\mathcal{C}_0 = \mathcal{B}_1$ gives a conservative estimate (upper bound) of performance that is nevertheless tighter than the upper bound provided by the uncoordinated performance. Coordinated scheduling outperforms uncoordinated scheduling as predicted by the analytical results providing a $5$ and $6$ dB gain with $N=10$, $20$, respectively, for an outage probability of $0.1$.

\subsection{Dependence on D2D Link Distance}
Since the concept of D2D communications relies on taking advantage of the proximity between communicating devices, it is of interest to examine performance as a function of D2D link distance $r_d$. Figure 3 shows the outage probability $\mathbb{P}(\textrm{SIR} < \theta_0)$ for SIR values $\theta_0 = -10$, $0$, $10$ dB, and $N=10$, as a function of $r_d$, under coordinated scheduling (simulation and analytical results) as well as under uncoordinated scheduling. It can be seen, that, in both cases, outage probability increases with $r_d$ and $\theta_0$. Coordinated scheduling always outperforms uncoordinated scheduling with most significant gains for values of $r_d$ up to about $1.2/(2\sqrt{\lambda_a})$. Note that the analytical results are very close to the simulation results for this range of $r_d$. Larger values of $r_d$ result in diminishing the gain offered by coordination, which is actually eliminated completely for very large $r_d$. This is due to D2D TXs and corresponding RXs residing in different cells, which is also the reason why analytical results become inaccurate for very large $r_d$ since Assumptions 1 and 4 no longer hold. However, this is not a serious shortcoming as this region of operation (very large $r_d$) is not of interest for D2D communications.

\begin{figure}
\centering
\resizebox{8.5cm}{!}{\includegraphics[trim = 10mm 5mm 10mm 10mm]{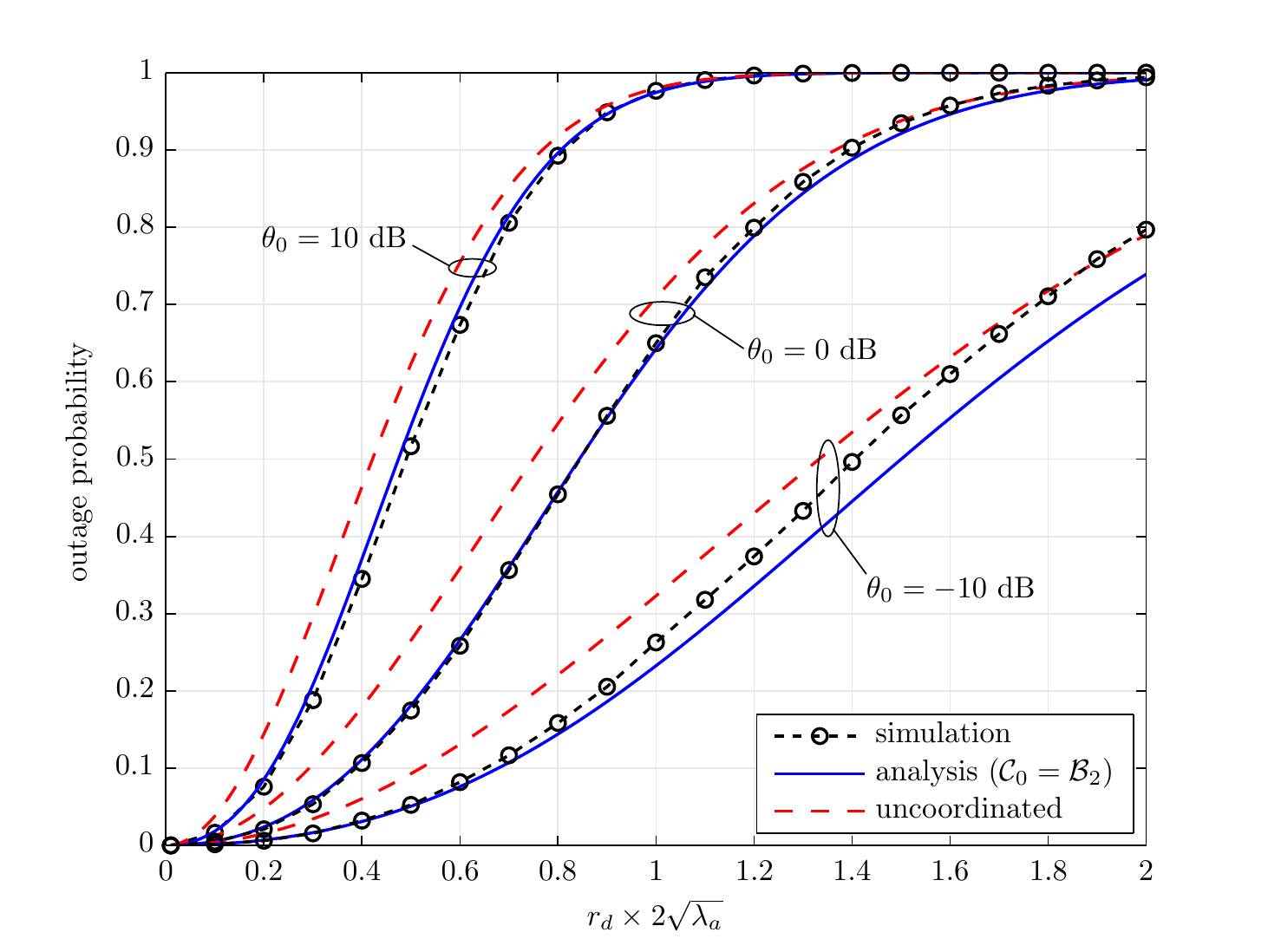}}
\caption{Outage probability as a function of $r_d$ ($\alpha = 4$, $\lambda_a=1$, $\lambda_d=10$, $N=10$).}
\end{figure}

\subsection{Benefits of D2D Communications}
Although SIR performance can be improved by increasing $N$, this comes at a cost of reduced bandwidth utilization, which suggests that $N$ must be optimized according to some rate-related criterion. To this end, the analytical expressions of Sec. III can be employed for an efficient numerical search of the optimal $N$. As an application, the following simple case study is considered that provides some insights on the benefits of D2D communications as overlay to the downlink of a cellular network. 

In particular, consider the presence of cellular RXs, i.e., RXs whose data are generated from sources that cannot allow for D2D communication, with locations modeled as an independent homogeneous PPP $\Phi_c$ of density $\lambda_c$. These RXs are served by their closest AP via a simple time-division-multiple-access (TDMA) scheduling scheme, with all APs in the system transmitting with the same power. The SIR threshold model for achieved rates is considered, i.e., a spectral efficiency of $\log(1+\theta_0)$ bits/Hz per channel use is achieved as long as the connection SIR is above a threshold $\theta_0$. 

Assuming a common SIR threshold for both cellular and D2D communications, the average user rate, taking into account the effects of TDMA for cellular users and $N>1$ for D2D communications, is defined as
\begin{equation} \label{av_rate}
R=\frac{\lambda_c}{\lambda_c+\lambda_d}(1-\eta)R_c + \frac{\lambda_d}{\lambda_c+\lambda_d}\eta R_d, \textrm{ (b/s/Hz),}
\end{equation}
where $R_c \triangleq \mathbb{E}(1/K_c)\mathbb{P}(\textrm{SIR}_c \geq \theta_0) \log(1+\theta_0)$, $R_d \triangleq \frac{1}{N}\mathbb{P}(\textrm{SIR}_d \geq \theta_0)\log(1+\theta_0)$, $K_c$ is the number of cellular RXs located within a random cell, $0<\eta<1$ is the portion of the downlink cellular bandwidth devoted for D2D communications, and $\textrm{SIR}_c$, $\textrm{SIR}_d$ stand for the SIR experienced by cellular and D2D RXs, respectively. The distribution of $K_c$ and $\textrm{SIR}_c$ have been exactly computed in \cite{Yu} and \cite{Andrews}, respectively. Therefore, $R$ is a function of $\eta$, $N$ and $r_d$ that can be optimized with the analytical formulas for $\textrm{SIR}_d$ of Sec. III. 

Figure 4 shows $R$ as a function of $r_d$ for  $\lambda_c = \lambda_d = 10$, $\lambda_a=1$, $\theta_0=0$ dB, and $\eta=\lambda_d/(\lambda_c+\lambda_d)$ (fair bandwidth partition). For each $r_d$, the optimal value of $N$ was obtained by a numerical search. For reference, performance when D2D communications are not supported is also shown, corresponding to the case $\lambda_c=20$ and $\eta=0$.\footnote{This implies that the cellular system is downlink limited with respect to user rate which is reasonable when the density of downlink users exceeds the density of uplink users.} For the coordinated case, performance obtained using (time-consuming) stochastic optimization based on simulation is also shown, matching very well the analytical results. It can be seen that introducing D2D communications to the system enhances the average user rate for values of $r_d$ up to about $80\%$ of the average distance from the closest AP, above which D2D communications perform worse than cellular and contribute negatively to the average user rate. This maximum distance suggests design guidelines, e.g., on the device-discovery training sequence/power, so as not to allow establishment of D2D links of distance greater than this threshold. It can also be seen that coordinated scheduling outperforms uncoordinated scheduling, and allows for about $9\%$ increased maximum allowed D2D link distance. The advantage of coordinated scheduling would be greater if reliability constraints, e.g., on SIR, were considered. Finally, performance with $N=1$ is also shown, which is much inferior both in terms of performance and maximum D2D link distance, indicating the importance of allowing for more than one SCs.

\begin{figure}
\centering
\resizebox{8.5cm}{!}{\includegraphics[trim = 10mm 5mm 10mm 10mm]{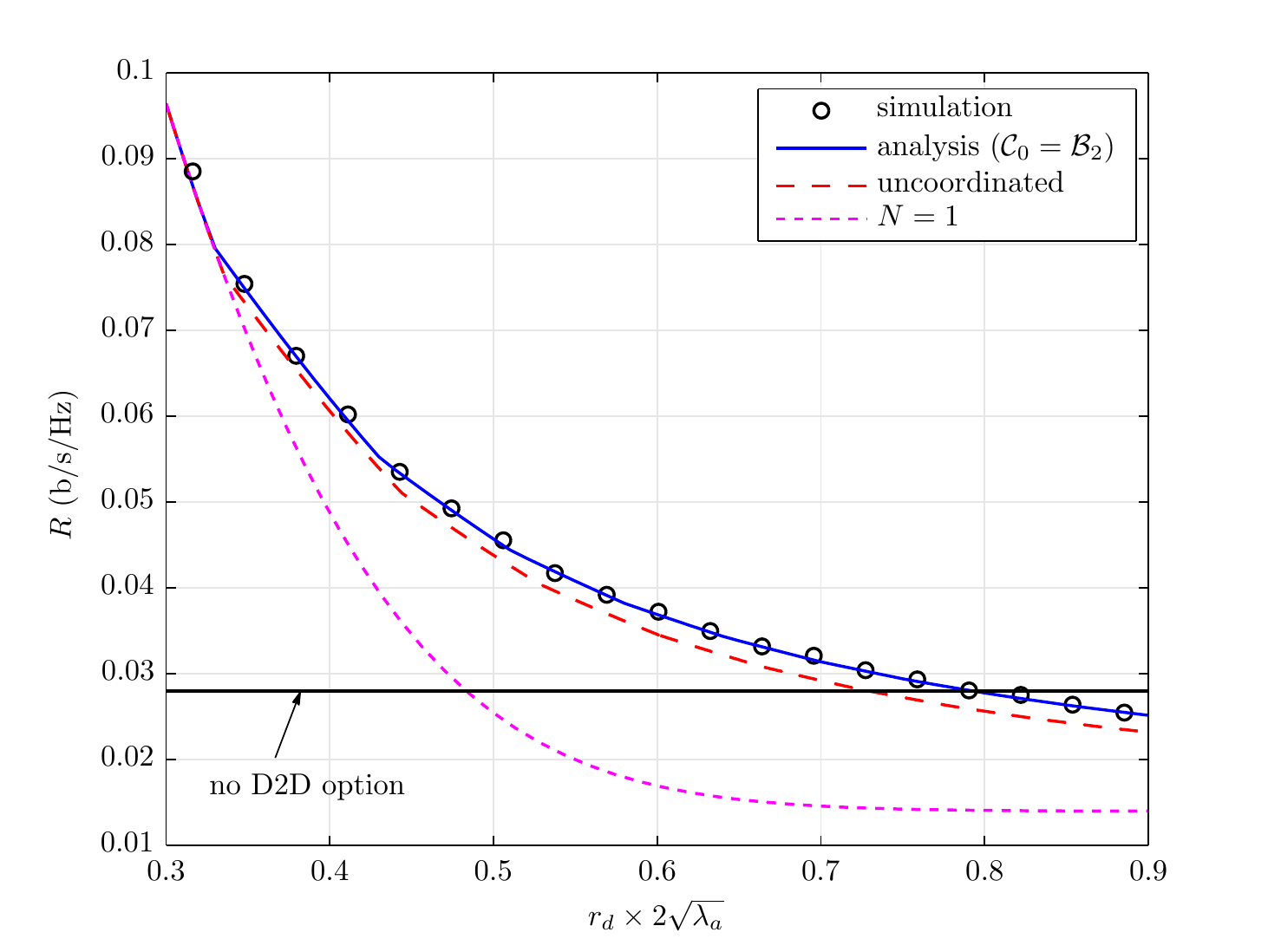}}
\caption{Average user rate as a function of D2D link distance with optimized $N$ ($\theta_0=0$ dB, $\alpha = 4$, $\lambda_a=1$, $\lambda_d=\lambda_c=10$).}
\end{figure}

\section{Conclusion}
This paper investigated the performance of overlay-inband D2D communications with coordinated scheduling directed by the cellular infrastructure. A simple scheduling scheme was assumed for which analytical expressions of the density of interferers as well as the SIR distribution were obtained, the latter validated by simulations. It was shown that coordination provides significant gains compared to an uncoordinated scheduling scheme, both in terms of SIR as well as average user rate for a system supporting cellular and overlay D2D communications. The analytical formulas can be utilized for system design analysis/optimization, e.g., for obtaining the maximum D2D link distance above which D2D communications are not beneficial compared to cellular.

\appendices

\section*{Acknowledgment}
This work has been performed in the context of the ART-COMP PE7(396)\textit{ ``Advanced Radio Access Techniques for Next Generation Cellular NetwOrks: The Multi-Site Coordination Paradigm''}, THALES-INTENTION and THALES-ENDECON research projects, within the framework of Operational Program ``Education and Lifelong earning'', co-financed by the European Social Fund (ESF) and the Greek State.

\end{document}